\documentclass{aa}

\usepackage[varg]{txfonts}

\usepackage{graphicx}
\usepackage{comment}

\usepackage{natbib}
\bibpunct{(}{)}{;}{a}{}{,}

\begin{document}

\title{Unveiling the formation of NGC 2915 with MUSE: \\ A counter-rotating stellar disk embedded in a disordered gaseous environment\thanks{Based on observations made with ESO Telescope at Paranal Observatory under program ID 094.B-0745(A)}}

\author{Yimeng Tang\inst{\ref{inst1},\ref{inst2},\ref{inst3}}
	\and Bojun Tao\inst{\ref{inst1},\ref{inst2}}
	\and Hong-Xin Zhang\thanks{Corresponding author}\inst{\ref{inst1},\ref{inst2}}
	\and Guangwen Chen\inst{\ref{inst1},\ref{inst2}}
	\and Yulong Gao\inst{\ref{inst4},\ref{inst5}}
	\and Zesen Lin\inst{\ref{inst1},\ref{inst2}}
	\and Yao Yao\inst{\ref{inst1},\ref{inst2}}
	\and Yong Shi\inst{\ref{inst4},\ref{inst5}}
	\and Xu Kong\inst{\ref{inst1},\ref{inst2}}
	}
\institute{Key Laboratory for Research in Galaxies and Cosmology, Department of Astronomy, University of Science and Technology of China, Hefei 230026, China\\ \email{ymtang@ucsc.edu, hzhang18@ustc.edu.cn} \label{inst1}
\and School of Astronomy and Space Science, University of Science and Technology of China, Hefei 230026, China\label{inst2}
\and Department of Astronomy and Astrophysics, University of California, Santa Cruz, 1156 High Street, Santa Cruz, CA 95064, USA\label{inst3}
\and School of Astronomy and Space Science, Nanjing University, Nanjing 210093, China\label{inst4} 
\and Key Laboratory of Modern Astronomy and Astrophysics (Nanjing University), Ministry of Education, Nanjing 210093, China\label{inst5}\\
}

\date{Received XXX / Accepted XXX}

\abstract{
NGC 2915 is a unique nearby galaxy that is classified as an isolated blue compact dwarf based on its optical appearance but has an extremely extended H {\sc i} gas disk with prominent Sd-type spiral arms. To unveil the starburst-triggering mystery of NGC 2915, we performed a comprehensive analysis of deep VLT/MUSE integral field spectroscopic observations that cover the star-forming region in the central kiloparsec of the galaxy. We find that episodes of bursty star formation have recurred in different locations throughout the central region, and the most recent one peaked around 50 Myr ago. The bursty star formation has significantly disturbed the kinematics of the ionized gas but not the neutral atomic gas, which implies that the two gas phases are largely spatially decoupled along the line of sight. No evidence for an active galactic nucleus is found based on the classical line-ratio diagnostic diagrams. The ionized gas metallicities have a positive radial gradient, which confirms the previous study based on several individual H {\sc ii} regions and may be attributed to both the stellar feedback-driven outflows and metal-poor gas inflow. Evidence for metal-poor gas infall or inflow includes discoveries of high-speed collisions between gas clouds of different metallicities, localized gas metallicity drops and unusually small metallicity differences between gas and stars. The central stellar disk appears to be counter-rotating with respect to the extended H {\sc i} disk, implying that the recent episodes of bursty star formation have been sustained by externally accreted gas.}


\titlerunning{Unveiling the Formation of NGC 2915 with MUSE}
\authorrunning{Tang et al.}
\maketitle

\section{Introduction}

Blue compact dwarfs (BCDs) are dwarf galaxies with compact optical appearance and H {\sc ii}-region-like spectra due to highly concentrated starburst activity \citep[e.g.,][]{Thuan1981,Hunter2004,Hunter2006}. They are generally metal-poor \citep[e.g.,][]{Kunth1988,Izotov1999}, and rich in gas which has a much more extended distribution than stars \citep[e.g.,][]{Brinks1988,Taylor1996,Putman1998,Pustilnik2001}. Such properties once led people to think that BCDs could be young systems that are experiencing their first episodes of star formation activity. However, later studies revealed the existence of extended ancient stellar populations in almost all of the known BCDs \citep[e.g.,][]{Loose1986,Kunth1988,Papaderos1996,Cairos2001,Cairos2017a,Cairos2017b}.\ Several scenarios have been put forward to understand the formation mechanisms of BCDs, including merger \citep[e.g.,][]{Bekki2008,Lelli2012,Zhang2020,Zhang2020b}, violent disk instabilities \citep[e.g.,][]{Elmegreen2012}, and cosmic web gas inflow\citep[e.g.,][]{Lopez-Sanchez2012,2014MNRAS.442.1830V}, etc., but it is unclear which mechanism is the most relevant one.

NGC 2915 is an extreme case of a BCD in the local Universe, which is also known as one of the closest BCDs ($M_{B}=-15.9$ mag; $D\sim4.1$ Mpc, \citealt{Meurer2003}), making it an ideal laboratory for understanding the evolution of this kind of galaxies. The H {\sc i} gas content of this galaxy is very high ($M_{\rm HI}$/$L_{B}\sim1.7~M_{\odot}/L_{\odot, B}$; $M_{\rm HI} \sim 4.5\times10^8 M_{\odot}$, \citealt{Elson2010}) and extremely extended in distribution ($\sim6\ R_{25}$, \citealt{Becker1988}), showing well-defined spiral structures that are not seen in optical images \citep{Meurer1996}. In the central region, there are two H {\sc i} gas clumps similar in mass and size \citep{Elson2010}, which were mistaken for a bar structure in low-resolution H{\sc i}~maps \citep{Meurer1996} for a long time. The H {\sc i} velocity field shows the regular rotation, and its rotation curve becomes flat at large radii \citep{Becker1988,Meurer1996,Elson2010}. Its dark matter content is extremely high ($M_{\rm tot}/L_B\sim140M_{\odot}/L_{\odot,B}$, \citealt{Elson2010}) and dominant at all radii. Also, the dark matter halo has a dense and compact core \citep{Bureau1999}, resulting in a very deep potential well in the center. Signs of a warp in the H {\sc i} disk are found \citep{Meurer1996}, indicating possible environmental interaction with a partner. A low-surface-brightness galaxy named SGC 0938.1-7623 \citep{Corwin1985,Karachentseva1998}, $\sim42'$ away from NGC 2915 (or projection distance is about 50 kpc by assuming the distance of NGC 2915), is the only candidate neighbor, without confirmation by either radial velocity or distance.

According to a study of multi-wavelength images of NGC 2915 by \cite{Meurer1994}, almost all the young stellar populations and H {\sc ii} regions are located near the center of the galaxy. \citet{Werk2010} studied five H {\sc ii} regions within the 1.2 $\times$ Holmberg radius and found a non-negative radial gradient of gas-phase metallicity. The apparent under-abundant inner disk and overabundant outer disk in metallicities indicate that there may be certain mixing mechanisms redistributing the heavy elements produced by star formation. In addition, \citet{Sersic1992} and \citet{Meurer1994} found that the optical surface brightness profiles of NGC 2915 are well described by an up-bending double-exponential form, with a sharp turning at a radius of $\sim$30 arcsec. The inner exponential disk, the so-called BCD part, mainly consists of the blue and young stellar population. whereas the outer exponential is red and old, similar to a dE galaxy. \citet{Meurer2003} discovered three metal-poor globular clusters based on HST images that cover a small portion of the outer disk, implying an abnormally high specific frequency of globular clusters for the whole system. Although being extremely rich in gas, star formation activity is barely detected in the extended outer H {\sc i} disk. The rotational shear may play a role in inhibiting the star formation there \citep{Meurer1996,Elson2012}.

\begin{figure}
\centering
\includegraphics[width=0.49\textwidth]{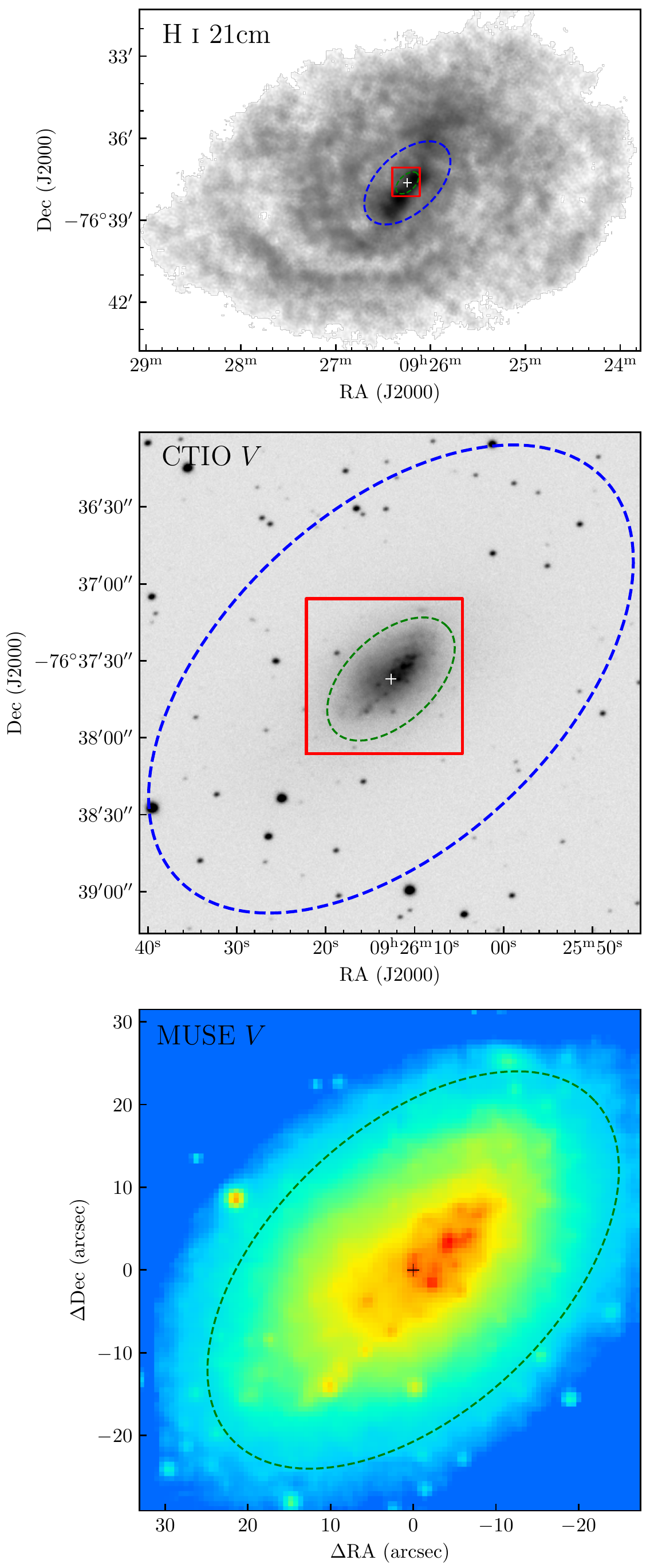}
\caption{From top to bottom: H {\sc i} intensity map of NGC 2915 obtained from the Australia Telescope Compact Array by \cite{Elson2010}, $V$-band image from the Cerro Tololo Inter-American Observatory's 1.5m telescope in the Spitzer Nearby Galaxies Survey \citep[SINGS,][]{Kennicutt2003}, and $V$-band image obtained by collapsing the wavelength dimension of the MUSE data cube. In all panels, the blue and green dashed ellipses represent the Holmberg radius and the break radius of the up-bending surface brightness profile of NGC 2915, respectively. The MUSE FoV is plotted as a red box, which measures $\sim$ 1.2$\times$1.2 kpc at the distance of NGC 2915. The galactic center is marked as a black cross and the same for all following plots.}
\label{fig:muse_fov}
\end{figure}

The triggering mechanism of the highly concentrated star formation in NGC 2915 is still unclear. In this study, taking advantage of the high-quality integral field spectroscopy from MUSE \citep{Bacon2010} on the Very Large Telescope (VLT), we present a study of the stellar population distributions and kinematics of NGC 2915, in an effort to shed light on the formation and evolution of this galaxy. The rest of this paper is structured as follows. Section \ref{sec:data} describes the method used in our data processing and analysis. The main results and discussion are given in Section \ref{sec:results}. A summary follows in Section \ref{sec:summary}.

\section{Observations and data reduction}\label{sec:data}

\subsection{MUSE data}

MUSE \citep{Bacon2010} is an integral-field spectrograph installed on the VLT of the European Southern Observatory at Cerro Paranal, Chile. The observations of NGC 2915 were taken in the wide-field mode (WFM) of MUSE, which covers a field of view (FoV) of $1'\times1'$, and has a spatial sampling of $0.2''\times0.2''$, a spectral resolution of $\sim2.5$ \AA{} (at 7000 \AA{}), and a spectral sampling of 1.25 \AA{} across the wavelength range from 4750 to 9350 \AA{}.

The MUSE observations of NGC 2915 (Figure \ref{fig:muse_fov}) were acquired over four nights in February 2015 (Program ID: 094.B-0745(A)) for a total exposure time of 12480 ($13\times960$) seconds on target, without separate sky exposures. The observations cover the central 1.2$\times$1.2 kpc of NGC 2915. We downloaded the raw science data and calibration data from the ESO Science Archive. The data were reduced with the MUSE data reduction pipeline (v2.8.3, \citealt{Weilbacher2020}) packaged in the ESO recipe execution tool (EsoRex, v3.13.3). We followed the standard procedure for calibration and reduction except for the sky subtraction. For observations without separate sky exposures, the default strategy adopted by the pipeline is to select the dimmest part of the white-light image (obtained by collapsing the data cube in the wavelength dimension) for sky background estimation. We adopt a more conservative (or stricter) strategy for background selection. In particular, we first select the faintest 10\% pixels in continuum and the faintest 10\% pixels in [O {\sc iii}]$\lambda$5007 emission, and then the common pixels from the two selections ($\sim 2.5$\% of the FoV) are used for background estimation. Finally, we refine the sky subtraction by applying the Zurich Atmosphere Purge (ZAP, \citealt{Soto2016}) principal component analysis algorithm. We note that the seeing conditions vary from exposure to exposure ($1.16''$ - $2.00''$), and the final combined data cube has a full width at half maximum (FWHM) of the point spread function (PSF) of $1.54''$, with little dependence on wavelength.

\subsection{HST images}
There exist multi-band archival HST/WFPC2 (Wide Field and Planetary Camera 2) images of NGC 2915, including the $U$ (F336W), $B$ (F439W), $V$ (F555W), and $I$ (F814W)-band images that were taken with exposure times of 4400, 2100, 600 and 600 seconds respectively (Program ID: 11987). The fully processed images of NGC 2915 are retrieved from the Hubble Legacy Archive, and are used in the current study for the following two purposes. First, to select a clean sample of star clusters from the relatively low-spatial-resolution MUSE data, we use the HST images to help remove contaminations from single stars (either foreground or within NGC 2915). Second, to have a better wavelength coverage for reliable stellar population modeling, the $U$ and $B$ band photometry are used jointly with the MUSE spectra for full-spectrum fitting (see Section \ref{sec:ppxf}). For the second purpose, we degrade the HST images to match the spatial resolution of MUSE data, using a Gaussian convolution kernel.

\section{Data analysis}
\subsection{2D spectrophotometric fitting}\label{sec:ppxf}
To prepare for resolved spectral fitting, we first rebin the MUSE datacube to have a $0.6''\times0.6''$ spaxel size. Then, all the spaxels that have ${\rm S/N}<3$ near 5500 \AA $\ $ or are contaminated by foreground stars or background galaxies are masked out. Finally, we use the Voronoi tessellation method (Vorbin, \citealt{Cappellari2003}) to adaptively rebin the spaxels to a minimum S/N of 50/\AA\ (near 5500\AA). The resultant spaxels with S/N $<$ 50/\AA\ are excluded from subsequent analysis. The same Voronoi binning mask is applied to the HST $U$, $B$, and $V$ band images for a joint fit of spectroscopy and photometry.

We use the Penalized Pixel Fitting (pPXF, \citealt{Cappellari2004,Cappellari2017}) software to perform a simultaneous spectral fit with stellar population models and Gaussian emission line templates to each adaptively rebinned spaxel. For the stellar population models, we use single stellar population (SSP) models generated from the Flexible Stellar Population Synthesis (FSPS, \citealt{Conroy2009}) code. These SSP models are built with the MILES stellar library, Padova+07 isochrones and the Kroupa initial mass function \citep{Kroupa2001}. We also include nebular continuum emission in our models. However, we note that whether or not the nebular continuum is included makes no significant difference to our results. Our SSP models cover 42 different ages ranging from 0.001 Gyr to 12.6 Gyr and 10 different metallicities from [M/H] $=$ $-1.98$ to $-0.20$. The model spectra are smoothed to match the spectral resolution of MUSE data.

By largely following the common practice of using pPXF in the literature \citep[e.g.,][]{Kacharov2018}, we developed the following procedure to perform a joint fit of the MUSE spectra and $UB$ broadband photometry, in order to obtain the stellar and gaseous velocity field, velocity dispersion, emission-line flux and star formation histories.

\begin{figure*}
\centering
\includegraphics[width=0.9\textwidth]{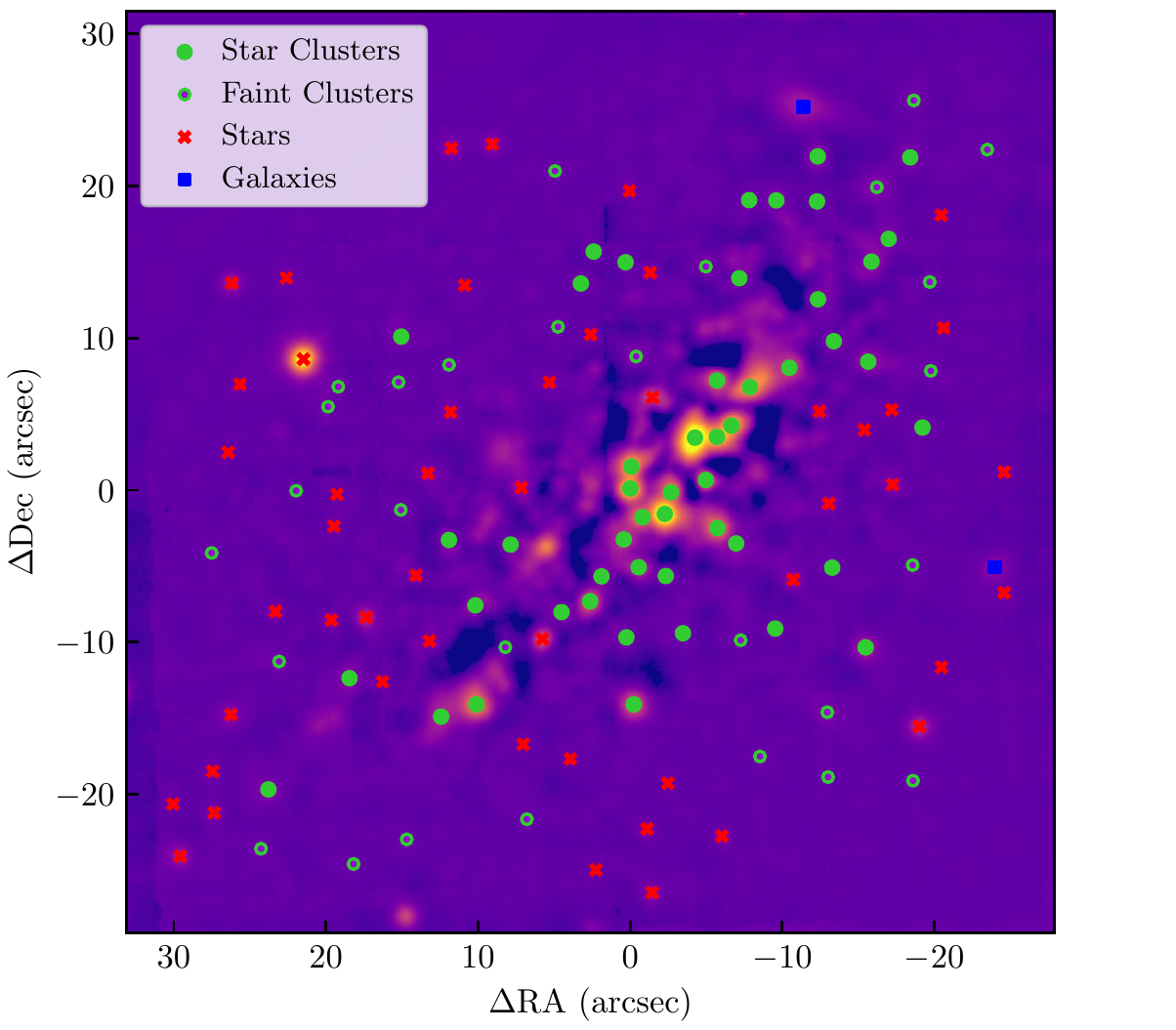}
\caption{Distribution of the compact sources detected based on the MUSE $V$-band image. Different types of sources are plotted with different symbols, as indicated in the legend. The background is the $V$-band image whose diffuse stellar light has been subtracted (see Section \ref{sec:star_cluster_method}). Bright star clusters ($V<22.8$ mag), faint star clusters ($V>22.8$ mag), stars, and background galaxies are shown as green filled circles, green open circles, red crosses, and blue squares.}
\label{fig:sources}
\end{figure*}

{\it Step 1}. We obtain the stellar radial velocity and velocity dispersion of each spaxel by performing a pPXF fitting without regularization. Emission lines are masked out at this point. In order to minimize effects such as template mismatch, imperfect sky subtraction, flux calibration and wavelength-dependent reddening, a combination of tenth-degree additive and multiplicative Legendre polynomials is used in the fitting. In the next steps, the stellar velocity and velocity dispersion are fixed to the values determined here.

{\it Step 2}. A pPXF fitting without regularization is performed again for each spaxel, using multiplicative tenth-degree Legendre polynomials only. As in step 1, emission lines are masked out at this point. We then use the \citet{Calzetti2000} reddening law to fit the multiplicative polynomials in order to determine the wavelength-dependent reddening. After dividing the multiplicative Legendre polynomials by the wavelength-dependent reddening, we obtain a residual continuum shape correction curve for each spaxel. These correction curves are median averaged over all spaxels to form a "master" correction curve, which is applied to all spaxels. We note that the typical correction factor is about 1\%. 

{\it Step 3}. The wavelength-dependent reddening curve determined above (and extrapolated to shorter wavelengths with the Calzetti law) for each spaxel is used to de-redden the corrected MUSE spectra and HST broadband photometry.

{\it Step 4}. Before performing a joint fit to the HST broadband photometry and MUSE spectra, we need to correct for the flux calibration difference (if any) between the two datasets. The HST $V$ band overlaps fully with the MUSE wavelength coverage. We therefore use the $V$ band to determine the overall flux calibration difference between HST photometry and MUSE spectra, and correct the HST $U$ and $B$ band photometry for this difference ($\lesssim$ 0.05 mag). We then rescale the full spectra+$UB$ noise array such that the corresponding pPXF fitting (without regularization) gives a minimum $\chi^2/{\rm DOF}$ = 1. Then, the Gaussian emission line templates are included in the fitting, and the relevant emission line parameters (i.e. flux, equivalent widths, velocity, and velocity dispersion) are measured.

{\it Step 5}. With all of the above corrections applied to the spectrum+$UB$, we perform pPXF fitting with third-order regularization \citep{Boecker2020} to iteratively find the best regularization parameter for each spaxel. The emission lines are masked out at this point. The best regularization parameter should increase the $\chi^2$ by $\Delta \chi^2\approx\sqrt{2N}$ compared to the unregularized best fitting, where $N$ is the number of wavelength pixels used in the fitting. The most likely nonparametric star formation histories are derived based on this final round of pPXF fitting.

\subsection{Detection and stellar population fitting of star clusters}\label{sec:star_cluster_method}

\begin{figure*}
\centering
\includegraphics[width=0.7\textwidth]{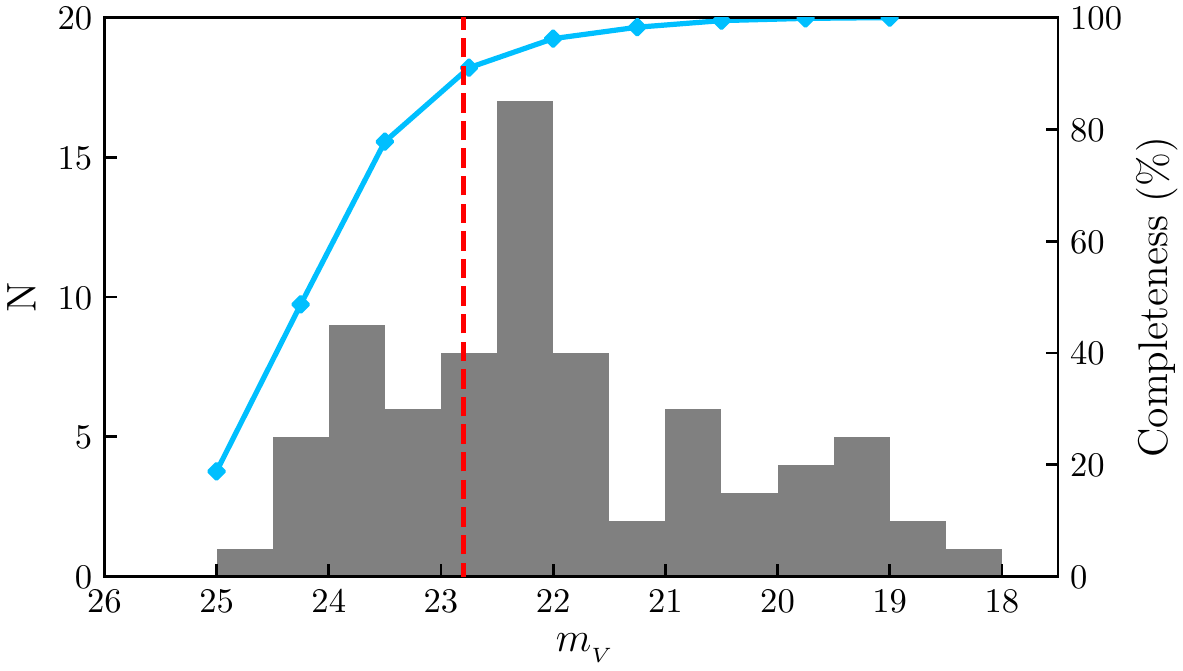}
\caption{$V$-band magnitude distribution of the star clusters in NGC 2915. The cyan curve indicates how completeness changes with source brightness. The red vertical dashed line marks the magnitude limit ($V=22.8$ mag) corresponding to the 90\% completeness. Only the star clusters brighter than this limit are used for our  analysis.}
\label{fig:completeness}
\end{figure*}

Star clusters are the best approximation to SSPs, and their distribution, although being affected by cluster dissolution processes, provides a relatively clean view of the star formation histories of their host galaxies. To maximize the cluster-detection efficiency in NGC 2915, we create a new data cube by combining the two MUSE exposures observed under the best seeing conditions (PSF FWHM = $1.16''$ and $1.18''$ respectively), and construct a $V$ band image by convolving each spectrum with a $V$ band filter curve and then collapsing the convolved data cube along the wavelength direction (Figure \ref{fig:sources}). In the rest of this section, we describe the cluster detection based on this $V$ band image, extraction of the spectra of the clusters, and fitting for stellar populations.

{\it Detection}. In order to better reveal star clusters, we use the \texttt{Background2D} task in the \texttt{Photutils} package to create a 2D map of the diffuse "background" light, with the parameter \texttt{box\_size} set to $10\times10$ pixels, and subtract this 2D background map from the $V$ band image. Next, the background-subtracted $V$ band image is used for cluster detection with the \texttt{DAOFIND} algorithm \citep{Stetson1987}. A total of 127 cluster candidates at least 5$\sigma$ above the local background are found. In order to estimate the completeness limit of our cluster detection, we adopt the artificial-star-test method, and find a 90\% completeness limit at $V\le22.8$ mag (Figure \ref{fig:completeness}).

{\it Spectro-photometric extraction}. A PSF-weighted aperture photometry is performed for the 127 cluster candidates in each wavelength slice of the original data cube. The PSF-weighted photometry is equivalent to the optimal extraction algorithm commonly used in long-slit spectroscopy. A Gaussian PSF model with ${\rm FWHM}=1.2''$ and a cutoff radius of $2.4''$ is adopted in the photometry. The local background for each cluster candidate is estimated with a circular annulus of inner radius of $2.2''$ and outer radius of $2.6''$, and the background spectrum is a median of all spaxels in the circular annulus. The same aperture photometry is also performed on the HST $U$ and $B$-band images.

{\it Sample cleaning}. The above-selected sample of cluster candidates is subject to contamination by single stars and background galaxies. We adopt two criteria to remove contaminants. First, the sources with radial velocities of greater than 150 km/s lower or higher than the systemic velocity of NGC 2915 ($\simeq465$ km/s) are regarded as foreground stars or background galaxies. Second, as typical star clusters at the distance of NGC 2915 are expected to be at least partially resolved on the HST images, sources with concentration indices ${\rm CI}_{2-3.5}$ measured on the HST white-light images (a combination of F336W, F439W, F555W, and F814W) smaller than 0.75 are regarded as single stars, where ${\rm CI}_{2-3.5}$ is the magnitude difference between an aperture of 2 pixels in radius and one of 3.5 pixels in radius. The ${\rm CI}_{2-3.5}$ criterion is based on the ${\rm CI}_{2-3.5}$ distribution of foreground stars confirmed with radial velocities in the NGC 2915 field. By applying the two criteria to the candidate sample, we are left with 78 confirmed star clusters, of which 49 are brighter than the 90\% completeness limit of $V=22.8$ mag. The following analysis is focused on these 49 clusters.

{\it Fitting with SSP Models}.
To estimate the ages and metallicities of the star clusters, we largely follow the fitting procedure elaborated in Section \ref{sec:ppxf} except that the star clusters are fitted as SSPs instead of combinations of multiple SSPs. We perform a joint fit of MUSE spectra and HST photometry with each SSP model and calculate the $\chi^2$ parameter. The most likely ages and metallicities are derived as the $\exp(-\chi^2/2)$-weighted median values, and the corresponding uncertainties correspond to the central 68\% of the cumulative distributions of $\exp(-\chi^2/2)$ for ages and metallicities respectively.

\section{Results}\label{sec:results}

\begin{figure*}
\centering
\includegraphics[width=1\textwidth]{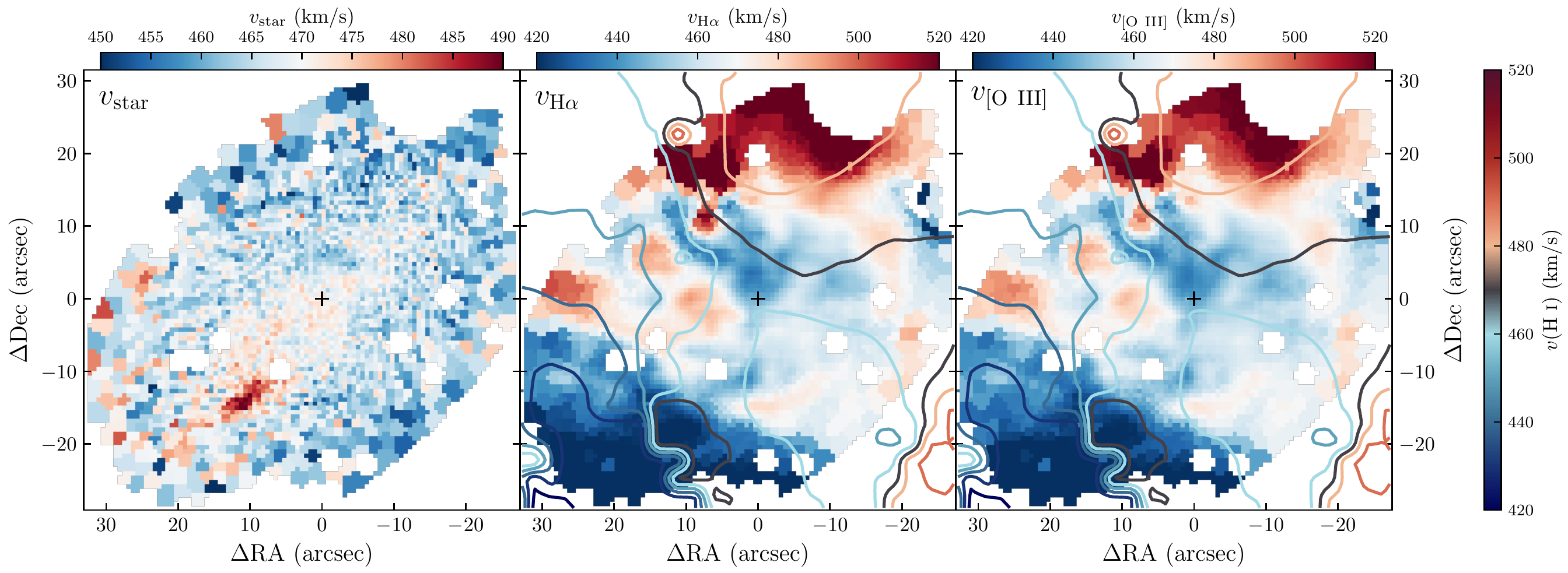}
\caption{Velocity field of the stars (left), H$\alpha$ emission line (middle), and [O {\sc iii}] $\lambda 5007$ emission line (right) of NGC 2915. Velocity field of the H {\sc i} 21-cm emission line \citep{Elson2010} is overplotted as contours.}
\label{fig:vmap}
\end{figure*}

\begin{figure}
\centering
\includegraphics[width=0.47\textwidth]{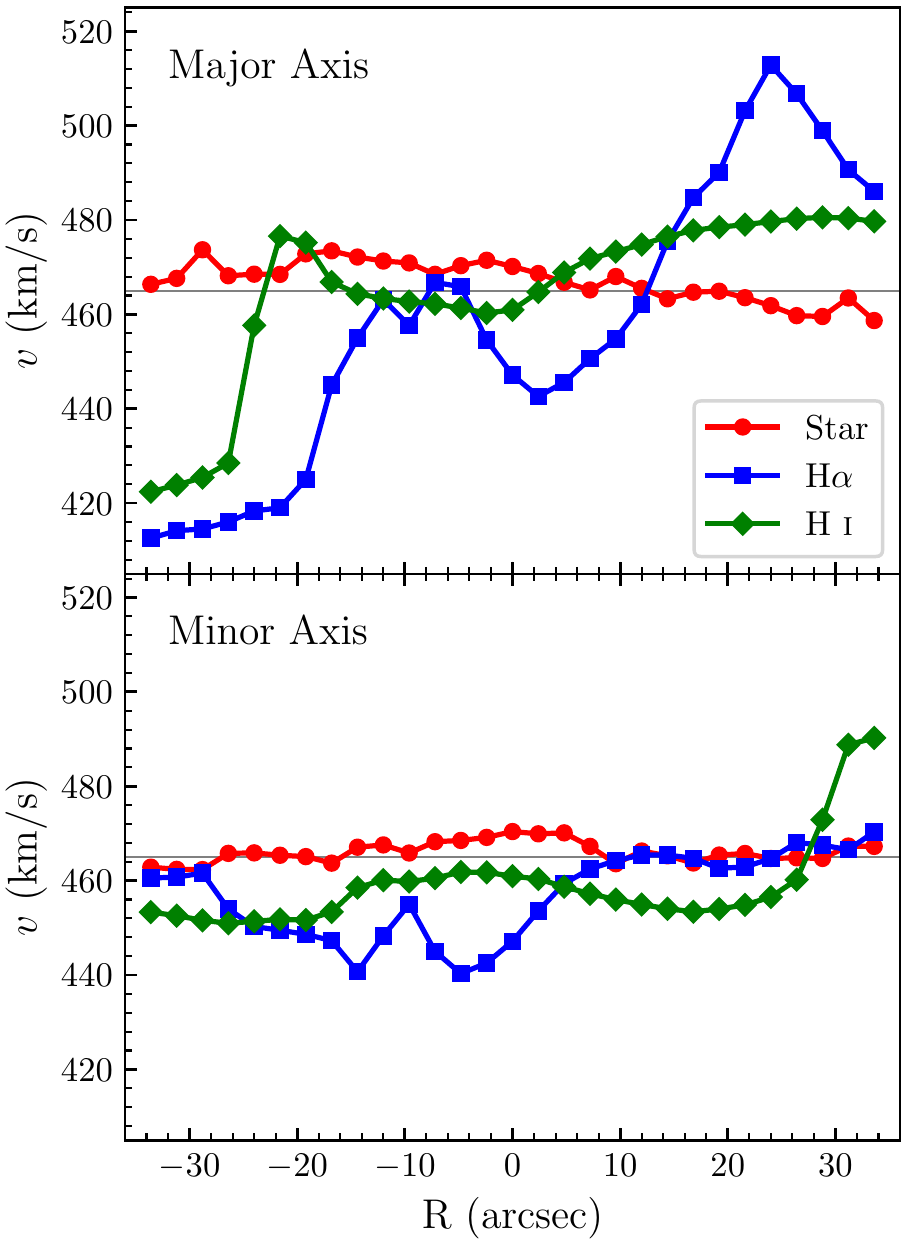}
\caption{
Radial variations of the average line-of-sight velocities along the photometric major (upper panel) and minor (lower panel) axes. A 8$''$-wide slit is used to extract the average velocities for both panels. The meanings of symbols of different shapes and colors are indicated in the upper panel.
}
\label{fig:velprofiles}
\end{figure}

\begin{figure*}
\centering
\includegraphics[width=1\textwidth]{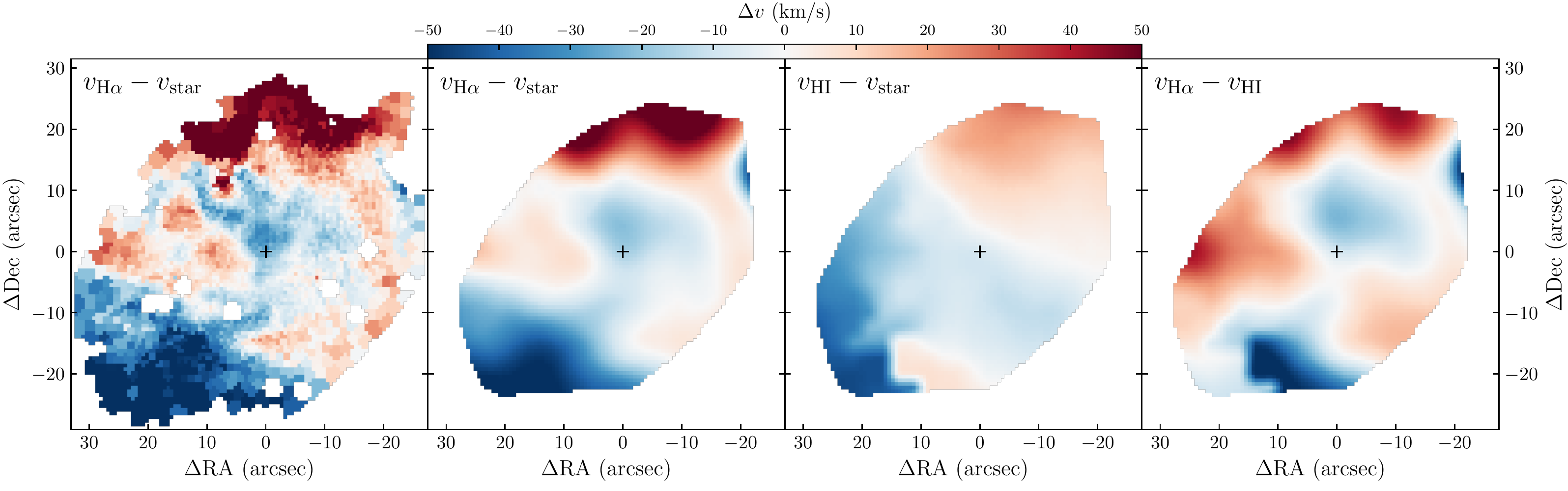}
\caption{
Velocity difference maps of the stellar and gaseous velocity fields. The left panel is the ($\upsilon_{\rm H\alpha}$$-$$\upsilon_{\rm star}$) map at the original spatial resolution of MUSE data, and the other three panels to the right are respectively the ($\upsilon_{\rm H\alpha}$$-$$\upsilon_{\rm star}$), ($\upsilon_{\rm H {\sc i}}$$-$$\upsilon_{\rm star}$) and ($\upsilon_{\rm H\alpha}$$-$$\upsilon_{\rm H {\sc i}}$) maps, where the MUSE data have been smoothed to match the spatial resolution (FWHM $\sim$ 18$''$) of H {\sc i} data.
}
\label{fig:deltavmap}
\end{figure*}

\begin{figure*}
\centering
\includegraphics[width=0.95\textwidth]{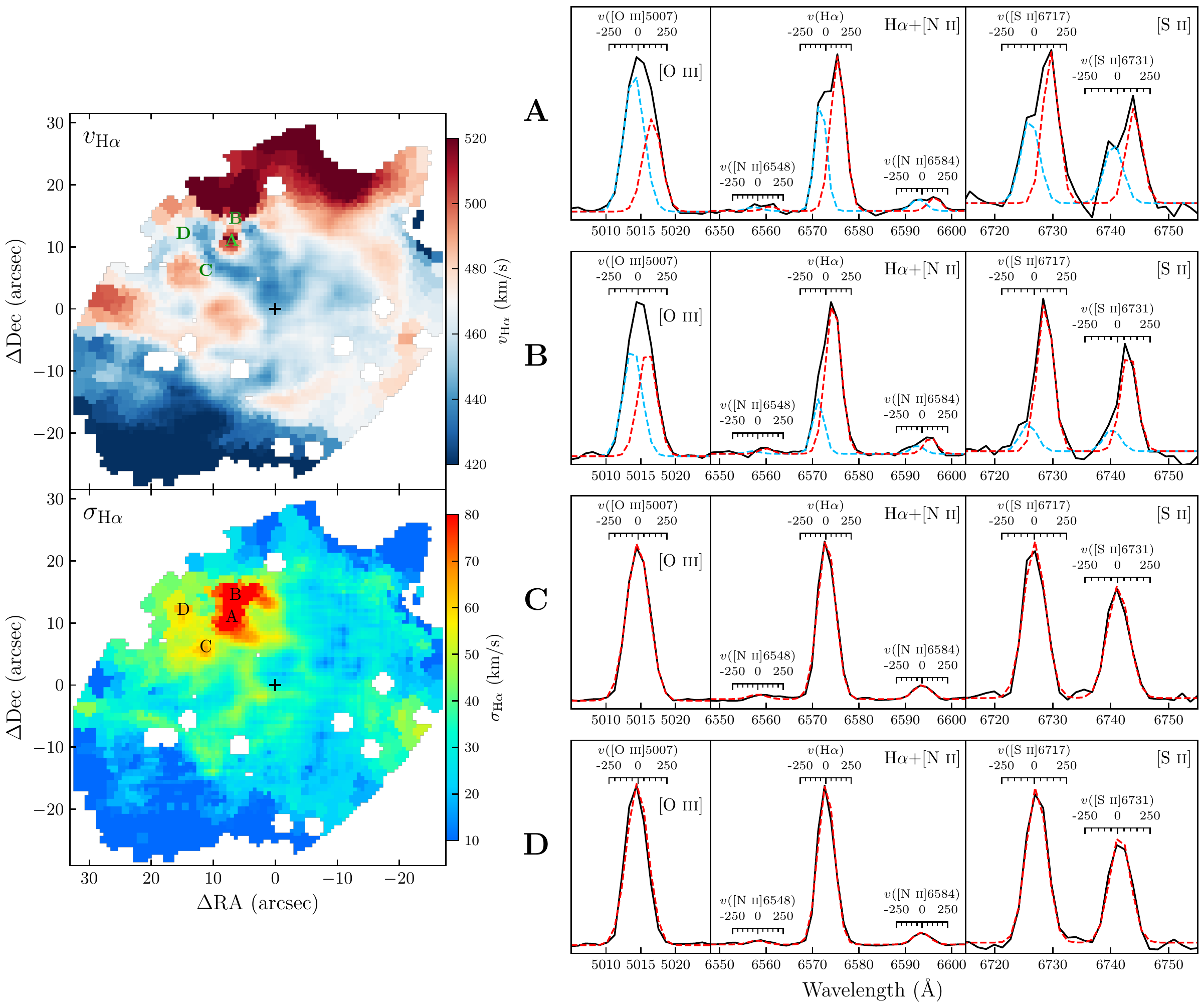}
\caption{{\em Left column}: Velocity field and velocity dispersion map of H$\alpha$ emission line. The velocity dispersion is determined by enforcing single Gaussian profile fitting. {\em Right column}: Emission line profiles of [O {\sc iii}]$\lambda$5007, H$\alpha$, [N {\sc ii}]$\lambda \lambda$6548,6584, and [S {\sc ii}]$\lambda \lambda$6717,6731 of the four regions with the highest velocity dispersion, as indicated in the left panels. The dashed curves overplotted on the line profiles represent the best fit with either single or double Gaussian components. The velocity scale (in km/s) for each spectral line profile is indicated in the right panels.}
\label{fig:vsigma}
\end{figure*}

\begin{figure*}
\centering
\includegraphics[width=1\textwidth]{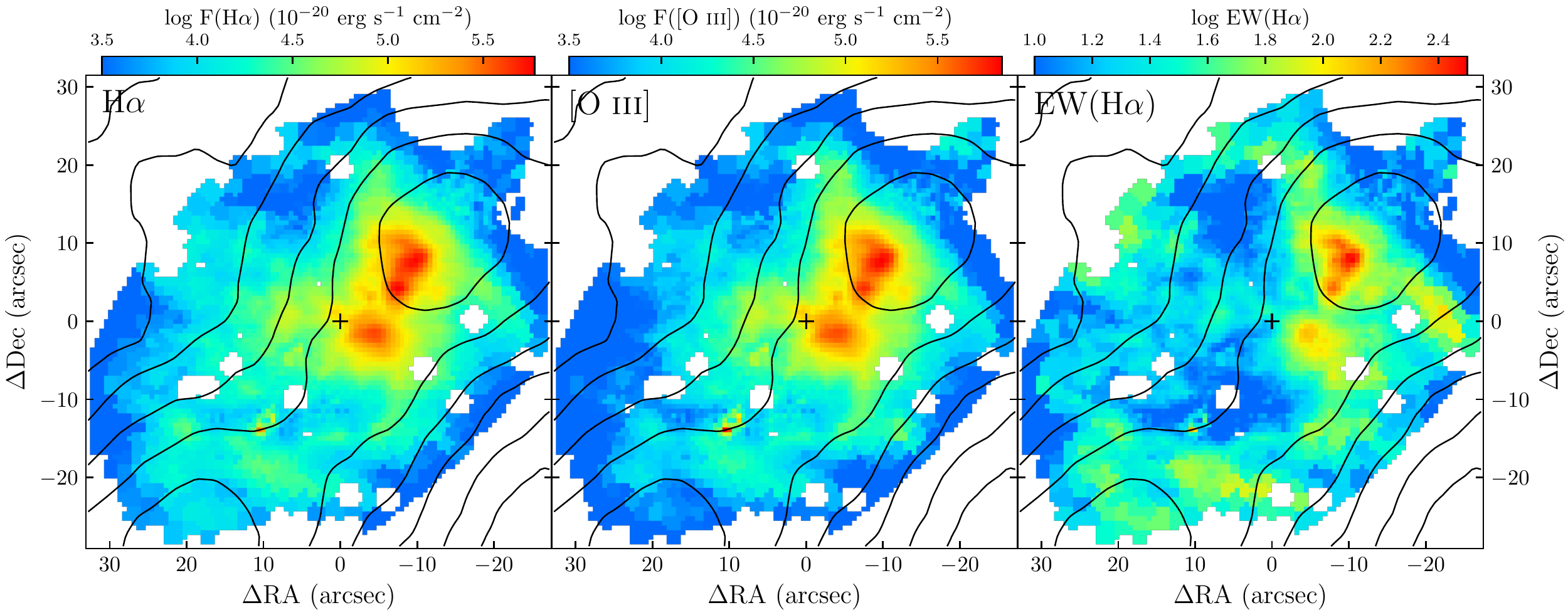}
\caption{Maps of H$\alpha$ intensity (left), [O {\sc iii}]$\lambda5007$ intensity (middle) and the equivalent width of H$\alpha$ (EW(H$\alpha$), right). The H {\sc i} 21-cm line intensity distribution \citep{Elson2010} is overplotted as contours.}
\label{fig:halpha_map}
\end{figure*}

\begin{figure*}
\centering
\includegraphics[width=1\textwidth]{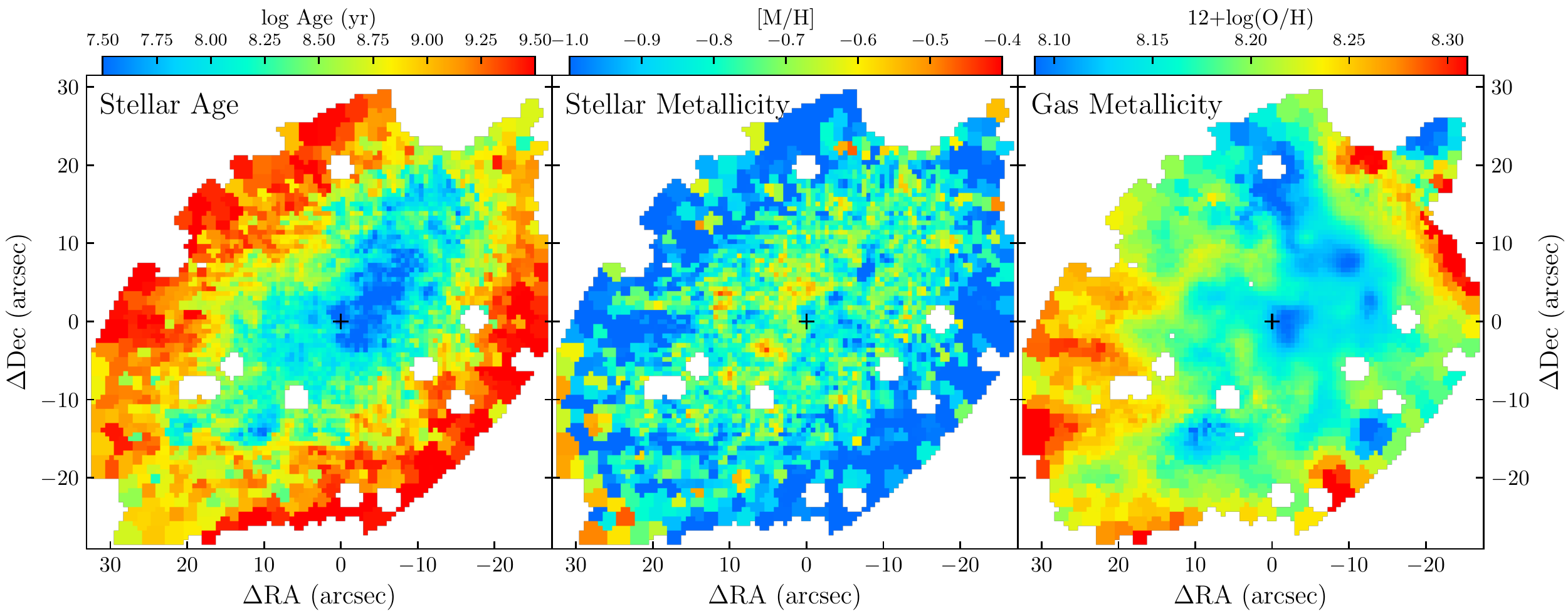}
\caption{Distributions of the light-weighted ages (left) and stellar metallicities (middle), and the metallicities of ionized gas calculated with the O3N2 method (right). The gas metallicities are calculated only for spaxels with ${\rm S/N}>3$ in H$\beta$, [O {\sc iii}]$\lambda5007$, H$\alpha$, and [N {\sc ii}]$\lambda6584$ emission lines.}
\label{fig:age_metal_map}
\end{figure*}

\subsection{Kinematics}

\subsubsection{The overall velocity gradients: a counter-rotating stellar disk} \label{sec:velfield}

Velocity fields of stars and ionized gas (as traced by H$\alpha$ and [O {\sc iii}]$\lambda$5007 emission lines) of NGC 2915 are shown in Figure \ref{fig:vmap}. The neutral H {\sc i} gas velocity field (synthesized beam $\sim$ 17$'' \times 18''$) from \citet{Elson2010} is overplotted as contours. As with the neutral gas, the ionized gas exhibits an obvious southeast--northwest velocity gradient that is approximately along the direction of the photometric major axis and of the H {\sc i} kinematic major axis. In contrast to the gas, the stars exhibit a much weaker velocity gradient that appears to be in an opposite sense to the gas.

To further quantify the radial-velocity variations, Figure \ref{fig:velprofiles} presents the radial variation of average velocities along 8$''$-wide slits that are aligned with either the photometric major (upper panel) or minor axes (lower panel). We note a negative southeast--northwest stellar velocity gradient, running from $\sim$ 472 km/s at $\sim-$20$''$  to $\sim$ 460 km/s at $\sim$ 30$''$ along the major axis. By adopting a photometric inclination angle of 55$^{\circ}$, the major-axis stellar velocity gradient corresponds to a maximum rotational velocity of $\sim$ 7 km/s within 30$''$. The stellar rotation is not just in the opposite direction to but also significantly lower than the rotational velocities ($\simeq$ 19 km/s at 30$''$) derived by tilted-ring fitting to the overall velocity field of the extended H {\sc i} disk \citep{Elson2010}. We checked the stellar velocity dispersion profile (not shown here) and found a nearly constant velocity dispersion of 39 km/s across the observed field, with the typical measurement error of $\sim$ 6 km/s and a scatter of $\sim$ 2 km/s along the major axis. The relatively low stellar rotational velocities may be attributed to a more important pressure support. In the stellar velocity map, we note that a small area near $\Delta$(RA) $\sim$ $10''$ and $\Delta$(Dec) $\sim$ $-15''$ has a significantly higher velocity than its surroundings. This high-velocity ``clump'' is associated (in projection) with several bright star clusters close to each other.

\cite{Elson2010} found that the H {\sc i} kinematics in the central region of NGC 2915 is best modeled by a highly inclined (85$^{\circ}$) disk together with radial expansion velocities of the order of $\sim$ 30 km/s (comparable to the rotational velocities). The ionized gas follows, albeit with significant local deviations, the overall velocity gradient of H {\sc i} gas, running from $\sim$ 420 km/s to $\sim$ 480 km/s along the major axis. The remarkably larger gaseous velocity gradient than the stellar ones (even if allowing for the possible inclination difference between the gaseous and stellar disks) confirms that the ionized and neutral atomic gas components are highly disturbed in the inner disk. It is noteworthy that the major-axis gaseous velocities decline abruptly beyond a radius of $\sim$ 20$''$ to the southeast. Based on the whole H {\sc i} velocity field of \cite{Elson2010}, this abrupt decline of major-axis velocities appears to be caused by a narrow gas stream projected across the inner and outer disks (see Figure 6 in \cite{Elson2010}).

\subsubsection{A complex gaseous kinematics: decoupling of ionized and neutral gas}\label{sec:disturbedgas}

To obtain a clear view of the disturbed gas kinematics, we subtract the stellar velocity field from the gaseous velocity fields and present the velocity difference maps in Figure \ref{fig:deltavmap}. In Figure \ref{fig:deltavmap}, the left panel is the map of line-of-sight velocity difference of H$\alpha$ ($\upsilon_{\rm H\alpha}$) and stars ($\upsilon_{\rm star}$) at their original spatial resolution, and the middle and right panels are respectively the maps of ($\upsilon_{\rm H\alpha}$$-$$\upsilon_{\rm star}$) and ($\upsilon_{\rm H {\sc I}}$$-$$\upsilon_{\rm star}$), where the MUSE maps have been smoothed to match the spatial resolution (FWHM $\sim$ 18$''$) of H {\sc i} data.

The convex iso-velocity contour lines of the H$\alpha$ emission to the northwest (middle panel of Figure \ref{fig:deltavmap}) are in contrast to the concave iso-velocity contour lines (as expected for a circular rotation) of the neutral H {\sc i} gas there, suggesting that the distribution of H {\sc i} gas is largely decoupled from the ionized gas along the line of sight. The velocity difference map of ($\upsilon_{\rm H\alpha}$$-$$\upsilon_{\rm H{\sc I}}$) shown in the right panel of Figure \ref{fig:deltavmap} more clearly illustrates the decoupling of ionized and neutral atomic gas. On the northwest side of NGC 2915, the major-axis of position--velocity curves of H$\alpha$ and H {\sc i} intersect at $R$ $\sim$ 15$''$ (Figure \ref{fig:velprofiles}), which roughly coincides with the peak of the most active star-forming region (Figure \ref{fig:halpha_map}). The major-axis variation of $\upsilon_{\rm H\alpha}$$-$$\upsilon_{\rm H{\sc I}}$ on the northwest side is approximately symmetric with respect to the peak of star formation at $R$ $\sim$ 15$''$ (Figure \ref{fig:velprofiles}). The spatial alignment between the H$\alpha$ kinematics and the active star-forming regions implies stellar feedback-driven outflows.

In addition to the above difference between the ionized and neutral gas toward the northwest, there are patches of redshifted regions to the east ($\Delta$(RA) $>$ 0; $\Delta$(Dec) $\sim$ $-5''-15''$) that are seen in the velocity maps of the ionized gas but not the neutral H {\sc i} gas. To explore the origin of these redshifted patches, Figure \ref{fig:vsigma} shows the H$\alpha$ velocity dispersion map derived from pPXF fitting (lower left panel). Several eastern regions with remarkably higher $\sigma_{{\rm H}\alpha}$ ($\gtrsim 70$ km/s) are indicated as A, B, C and D. The right panels of Figure \ref{fig:vsigma} show the line profiles, together with Gaussian profile fitting, of several nebular emission lines of the four indicated regions. For regions A and B which have the highest velocity dispersion, we find double-component emission line profiles. The two kinematic subcomponents have velocity differences of $\sim 160$ km/s, and none of them (409 and 579 km/s for the two subcomponents of A, and 387 and 534 km/s for B) are close to the expected disk velocities ($\sim$ 460 km/s), indicating either expanding super-bubbles or colliding gas clouds (e.g., \citealt{Nath2013,Sharma2014}). Nevertheless, the redshifted subcomponents have lower $\log$([N {\sc ii}]$\lambda \lambda$6584/H$\alpha$) values than the blueshifted subcomponents ($0.25\pm0.12$ dex lower for A, $0.31\pm0.42$ dex for B), suggesting lower metallicities for the redshifted subcomponents. Therefore, the double-component line profiles are best explained by cloud--cloud collisions. None of the colliding clouds is kinematically associated with the disk. These cloud--cloud collisions reflect ongoing gas infall toward the disk from opposite directions.

A comparison of the velocity dispersion map and the H$\alpha$ equivalent width map (Figure \ref{fig:halpha_map}) suggests that the local star formation in these high-$\sigma_{{\rm H}\alpha}$ areas is significantly lower than its surroundings. This is probably caused by high-speed cloud--cloud collisions \citep{Maeda2021}. For regions C and D, the line profiles are well fitted by single Gaussian functions, without the need of second Gaussian components. A local suppression of star formation is seen (albeit over a much smaller area than that for regions A and B) in region C but not in region D.

\subsection{Star formation, age and metallicities}\label{sec:csfr}

To explore the spatial distribution of the current star formation activities, Figure \ref{fig:halpha_map} presents the H$\alpha$ and [O {\sc iii}]$\lambda5007$ line intensity maps and the H$\alpha$ equivalent width EW(H$\alpha$) maps. Regions with bright emission lines are the sites with active current star formation. It is clear that the current star formation is concentrated towards the northwest side of the galaxy center, and the most intense star formation ($\Delta$(RA) $\sim$ $-10''$; $\Delta$(Dec) $\sim 8''$) appears to be associated with the northwest peak of H {\sc i} column densities.

\begin{figure}
\centering
\includegraphics[width=0.49\textwidth]{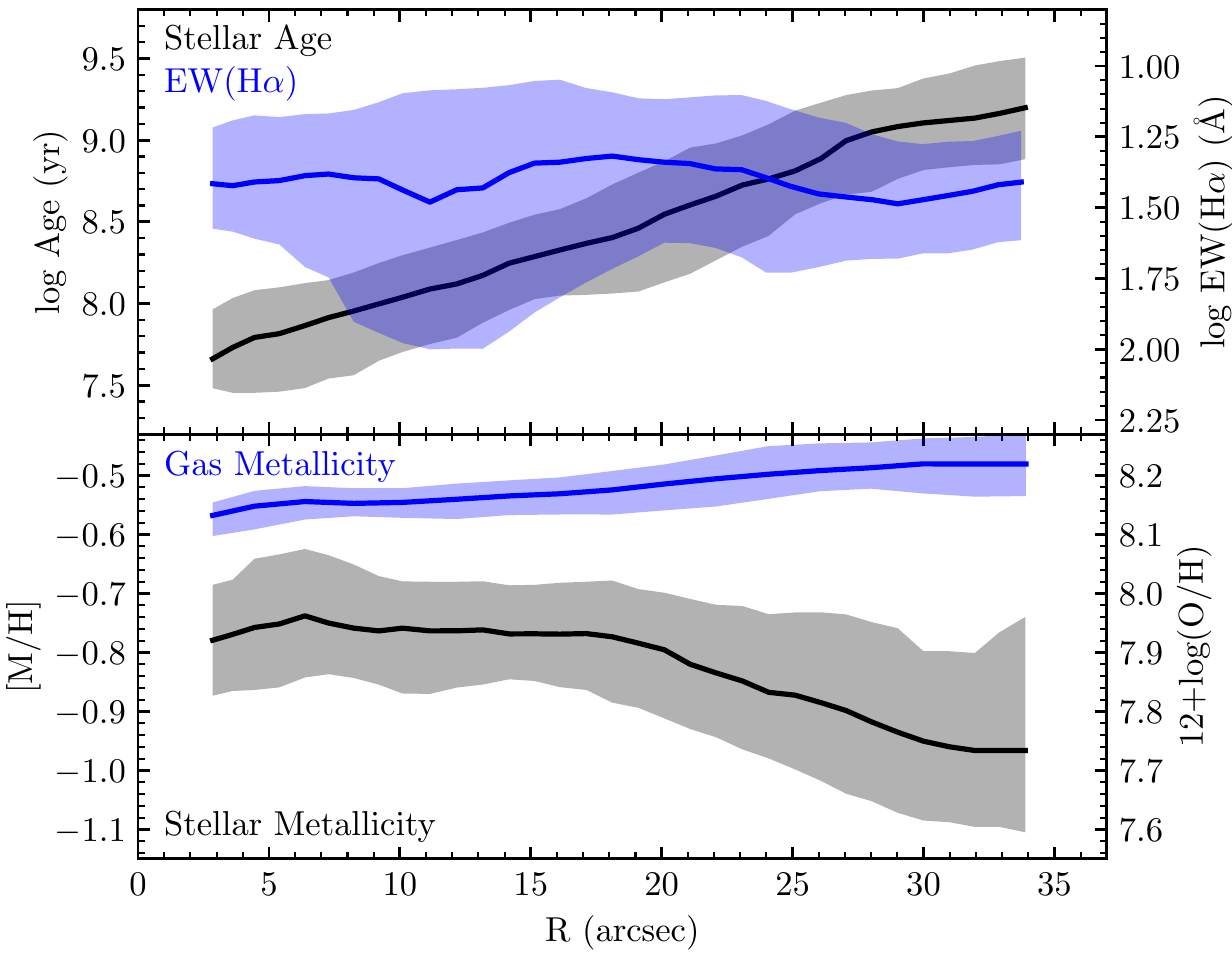}
\caption{Radial profiles of light-weighted stellar age and EW(H$\alpha$) (upper panel), light-weighted stellar metallicity and nebular gas oxygen abundances (lower panel). Elliptical annuli are used to extract these profiles. The lower and upper edges of the error shades correspond to the 16th and 84th percentiles of the azimuthal distribution of each plotted parameter.}
\label{fig:age_metal_profile}
\end{figure}

The light-weighted ages, stellar metallicities and ionized gas (oxygen) metal abundances are shown in Figure \ref{fig:age_metal_map}, and the azimuthally averaged radial profiles of these parameters are presented in Figure \ref{fig:age_metal_profile}. The gas metallicities are estimated with the O3N2 method \citep{Alloin1979}, which was recently found to give reliable metallicity estimations even in the low H$\alpha$ surface brightness regions where significant contribution from diffuse ionized gas (DIG) may be expected \citep{Kumari2019,ValeAsari2019}. Specifically, we adopt the formula derived by \citet{Marino2013}:
$$12+{\rm log(O/H)}=8.505-0.221\times{\rm O3N2}$$
where $${\rm O3N2}\equiv\log\left(\frac{[{\rm O\ III}]\lambda5007}{{\rm H}\beta}\times\frac{{\rm H}\alpha}{[{\rm N\ II}]\lambda6584}\right)$$

\begin{figure}
\centering
\includegraphics[width=0.4\textwidth]{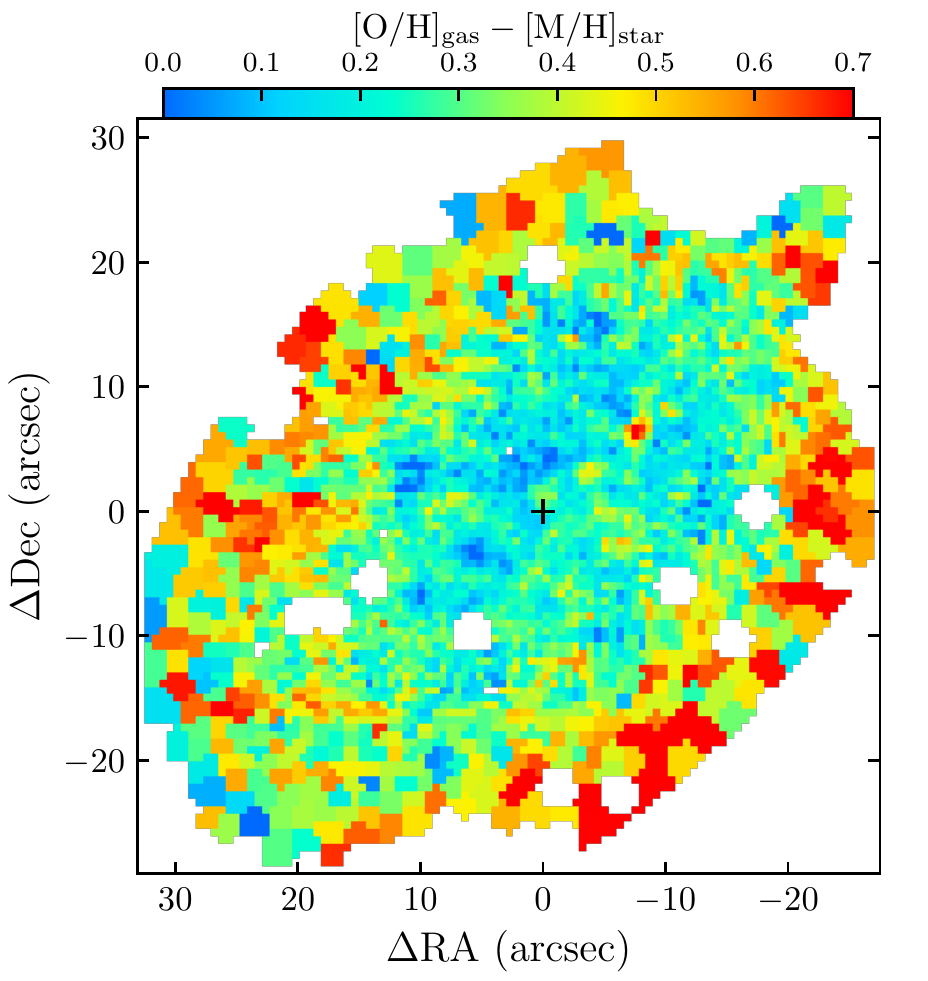}
\caption{Metallicity difference map of the ionized gas and stars. [O/H]$_{\rm gas}$ is defined as $12+\log ({\rm O/H}) - 8.69$, where 8.69 is the solar oxygen abundance \citep{Prieto2001}.}
\label{fig:metaldiffmap}
\end{figure}

The regions with the youngest stellar ages are located in the northwest quadrant and roughly along (albeit with an obvious offset) the photometric major axis. This is in agreement with the distribution of H$\alpha$ and [O {\sc iii}]$\lambda5007$ emission lines. The youngest light-weighted stellar ages are $\gtrsim$ 30 Myr, which is in line with the post-starburst nature of NGC 2915. The azimuthally averaged stellar ages steadily increase with galactocentric radii, in agreement with what is usually found in dwarf galaxies. For the stellar metallicity distribution, the azimuthally averaged profile is nearly flat at R $\lesssim$ 20$''$, beyond which the average metallicities decrease significantly with radius, with a negative gradient of $\simeq$ $-$0.73 dex/kpc. There is no significant azimuthal variation of stellar metallicities at a given galactocentric radius (Figure \ref{fig:age_metal_map}). The metallicity difference map of the gas and stars is shown in Figure \ref{fig:metaldiffmap}. 

The most active star-forming sites have lower gas metallicities than their immediate surroundings (right panel of Figure \ref{fig:age_metal_map}). This suggests that recent metal-poor gas inflow has played a role in triggering the active star formation \citep[e.g.,][]{Kumari2017}. The radial profile of average gas metallicities appears to be flatter than the stellar metallicity profile, but it has an overall positive gradient of $\sim$ 0.14 dex/kpc. This positive metallicity gradient confirms the finding of $\sim$ 0.12 dex/kpc in \cite{Werk2010} based on spectroscopy of five H {\sc ii} regions in NGC 2915.

A positive gas metallicity gradient is not expected for a galaxy whose metal production, mixing and loss are in equilibrium \citep{Sharda2021}. \cite{Werk2010} discussed three possible scenarios to explain the positive gradient: metal mixing across the disk, supernova-driven blowout and fallback, and galaxy interactions. Our kinematic analysis in Section \ref{sec:disturbedgas} suggests significant stellar feedback-driven outflows. Metal-enriched gas ejected by the feedback may eventually fall back to the disk, with larger angular momentum and thus larger galactocentric distances than before \citep[e.g.,][]{Brook2012}, contributing to a less negative or even positive metallicity gradient. Moreover, in addition to the clear evidence for gas infall (e.g., Section \ref{sec:disturbedgas}), signature of gas inflow is implied by the localized gas metallicity drops (right panel of Figure \ref{fig:age_metal_map}). Therefore, metal-poor gas infall or inflow has also played a role in shaping the gas-phase metallicity gradient. We note that the logarithmic differences between gas-phase and stellar metallicities (both normalized to the solar units; Figures \ref{fig:age_metal_profile} and \ref{fig:metaldiffmap}) are $\sim$ 0.2 dex at small radii ($<$ 20$''$) and $\sim$ 0.3-0.5 dex at large radii.\ These differences are $>$ 0.5 dex smaller than the typical values found in low-mass galaxies \citep[e.g.,][]{Lian2018}, implying a strong impact from either metal outflow or gas inflow.

\begin{figure}
\centering
\includegraphics[width=0.49\textwidth]{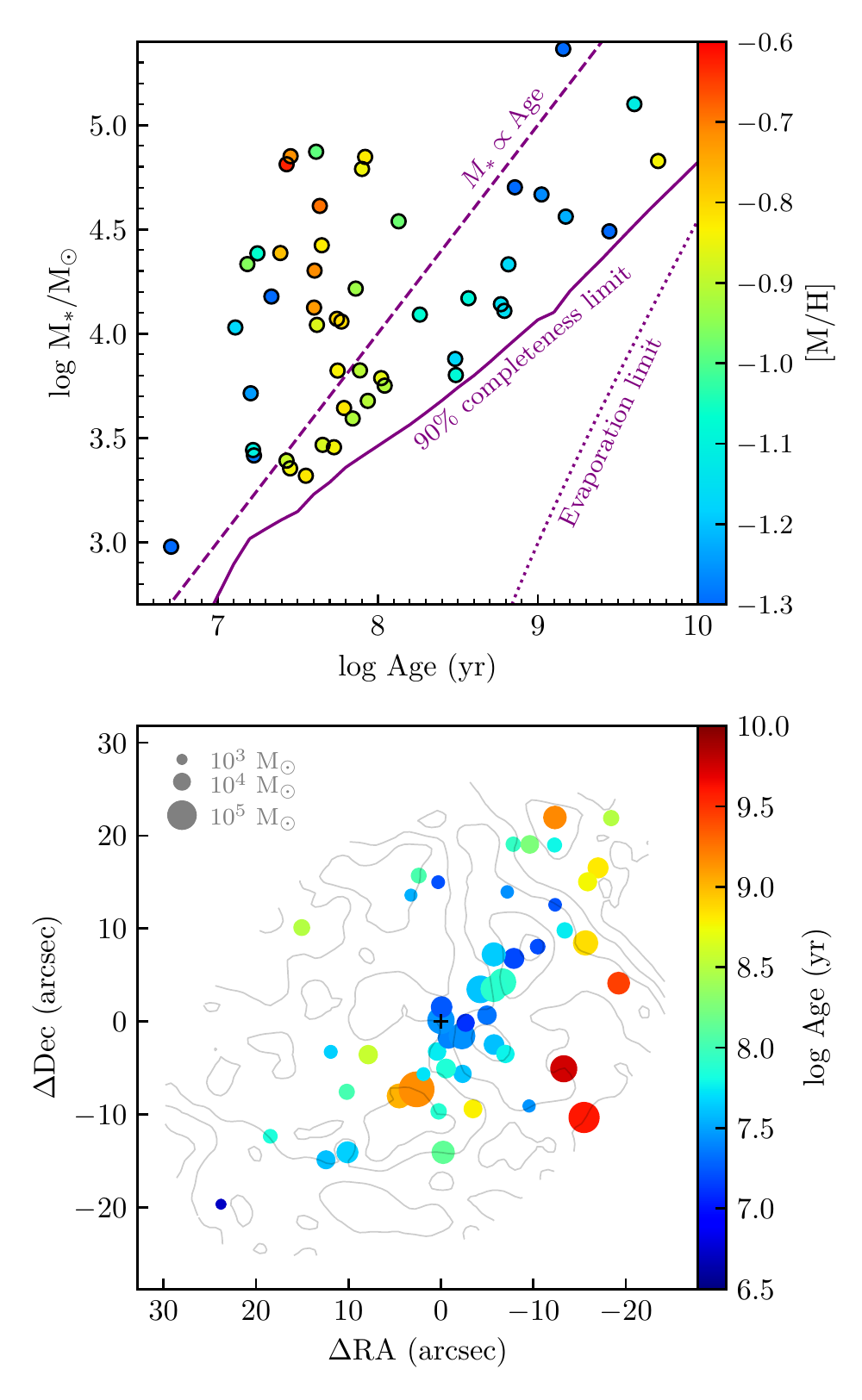}
\caption{Star cluster distributions of NGC 2915. {\em Top}: Distribution of star clusters in the mass--age plane. The plotted mass is the mass at birth. Each circle is a star cluster color-coded by its metallicity. The purple solid curve is the 90\% mass completeness limit converted from the $V=22.8$ mag brightness limit by using age-dependent mass-to-light ratios. The purple dashed line is plotted as a guide for the expected relation between maximum cluster mass and age by assuming a fiducial cluster mass function and constant cluster formation history. The dynamical evaporation limit is derived using a galactocentric radius of 0.47 kpc (24 arcsec) and a rotation velocity of 15 km/s, as appropriate for the inner disk of NGC 2915. {\em Bottom}: Spatial distribution of star clusters color-coded by the stellar age. The filled circles are plotted with sizes proportional to the fourth root of stellar mass. The EW(H$\alpha$) distribution is overplotted as contours.}
\label{fig:cluster_m_age}
\end{figure}

\subsection{Star cluster distribution}\label{sec:sc_history}

The age--mass distribution and spatial distribution of star clusters above the 90\% completeness detection limit are shown in Figure \ref{fig:cluster_m_age}. The maximum star cluster mass per logarithmic time interval is expected to increase with age ($M_{max} \propto {\rm Age}$ for a cluster mass function with a fiducial power-law slope of $-2$; \citealt[][]{Krumholz2019}) if the cluster formation rate is constant with time. A comparison with the $M_{max} \propto {\rm Age}$ relation implies that the age--mass distribution of the star clusters in NGC 2915 is inconsistent with a constant cluster formation history. There is an obvious enhancement of cluster formation at $\sim$ 10 $-$ 100 Myr ago. In addition, the most massive cluster is $\sim$ 2 Gyr old, implying a lower cluster formation rate at earlier times (more than a couple of gigayears). The long-term dynamical dissolution effect \citep{Baumgardt2003} on our clusters is not important above the 90\% completeness limit, as indicated in Figure \ref{fig:cluster_m_age}. We note that, although a quantitative understanding of the observed age--mass distribution of star clusters may be hindered by various very uncertain disruption and fading effects \citep[e.g.,][]{Elmegreen2010}, the features for NGC 2915 mentioned above cannot be easily explained by any known disruption mechanism.

Spatial distribution of the clusters is highly asymmetrical about the major axis (lower panel of Figure \ref{fig:cluster_m_age}), with more clusters in the southwestern half of the galaxy. The spatial concentration of clusters largely coincides with the active star-forming area (Section \ref{sec:csfr}). With that said, the region with the highest H$\alpha$ equivalent width ($\Delta$(RA) $\sim$ $-10''$; $\Delta$(Dec) $\sim$ 8$''$) has lower spatial concentration of clusters than regions at smaller galactocentric distances along the major axis, suggesting that the most active star formation very recently propagated to the current site from smaller galactocentric distances. Finally, we note that the two most massive clusters lie outside the active star-forming area.

\begin{figure*}
\centering
\includegraphics[width=1\textwidth]{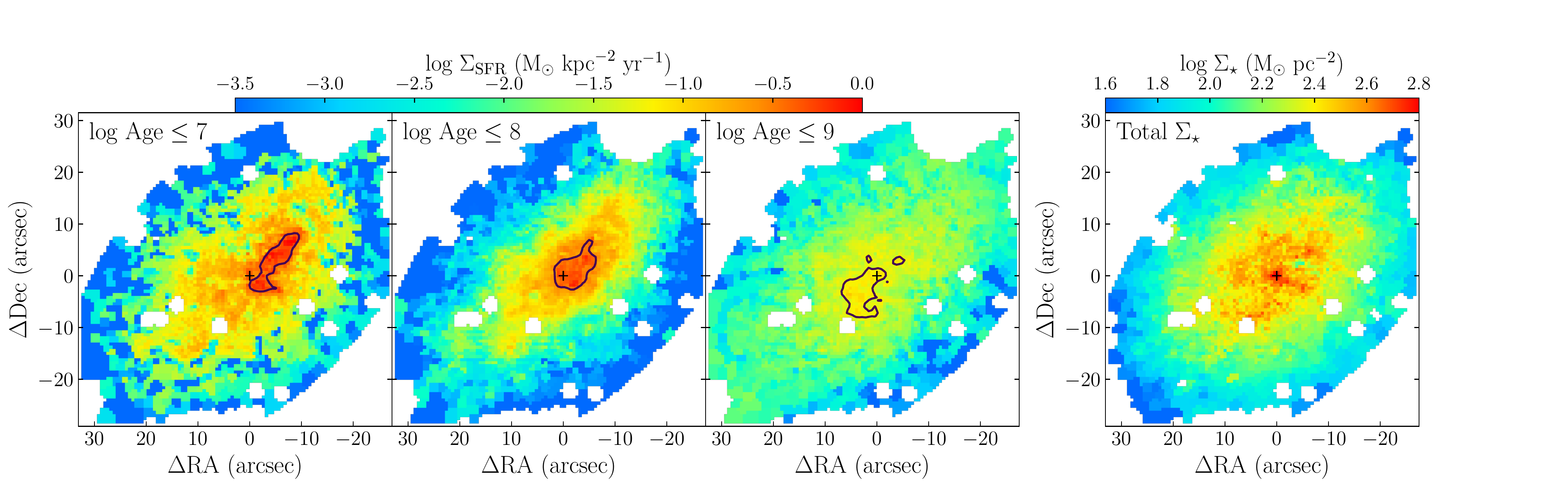}
\caption{Spatial distribution of star formation rate averaged over different timescales. {\em Left three panels}: Surface density distributions of star formation rate averaged over the last $10^7$, $10^8$ and $10^9$ yr (i.e., the birth mass of stars younger than the age limits divided by the age limits), respectively. The black contours overplotted in the left three panels correspond to 50\% of the maximum $\Sigma_{\rm SFR}$. {\em Rightmost panel}: Surface density distribution of stellar mass.}
\label{fig:sfh_2dmap}
\end{figure*}

\subsection{Star formation history} \label{sec:sfh}

\subsubsection{Spatially resolved star formation histories}

Spatially resolved star formation histories derived from pPXF fitting are presented in Figure \ref{fig:sfh_2dmap}. In order to alleviate the potential degeneracies between the estimate of star formation rate in adjacent age bins, we analyze the cumulatively averaged star formation histories. Surface density distribution of star formation rate or stellar mass averaged in cumulatively broader age bins (i.e. $\leq$ $10^7$ yr, $\leq$ $10^8$ yr, $\leq$ $10^9$ yr and throughout the whole lifetime) are shown in different panels of Figure \ref{fig:sfh_2dmap}.

We perform exponential radial profile fitting to the four maps shown in Figure \ref{fig:sfh_2dmap}, and find scale lengths of 3.9, 4.5, 11.1, and 18.4 arcsec respectively for the four age intervals. This means younger stellar populations are more spatially confined towards the central region of NGC 2915, which implies that either the star formation activities have become more and more centrally concentrated over time or older stars have had more time to migrate or disperse to larger radii. Nevertheless, the spatial distribution of star formation is not uniform along the azimuthal direction, as indicated by the iso-$\Sigma_{\rm SFR}$ contours in Figure \ref{fig:sfh_2dmap}. When averaged over the past $10^9$ yr, the southeastern side of NGC 2915 appears to have higher $\Sigma_{\rm SFR}$ than the other side. Over the past $10^8$ yr, the northwestern side has significantly stronger $\Sigma_{\rm SFR}$ than the southeastern side, as opposed to the average over $10^9$ yr timescales. In the last $10^7$ yr, star formation has continued to be more active in the northwestern side of the galaxy, and at the same time has been significantly enhanced along a $\sim$ 5$''$-wide ($\sim$ 0.1 kpc) stripe that is parallel to the direction of the major axis. The linear correlation of the stellar line-of-sight velocities with the radius shown in Figure \ref{fig:velprofiles} suggests a solid-body rotating stellar disk, as with the solid-body H {\sc i} rotation curve within the central $\sim$ 3 kpc \citep{Elson2010}, so the above-mentioned azimuthal asymmetry is (in a relative sense) not subject to significant rotational mixing.

\subsubsection{Global star formation history}

\begin{figure}
\centering
\includegraphics[width=0.468\textwidth]{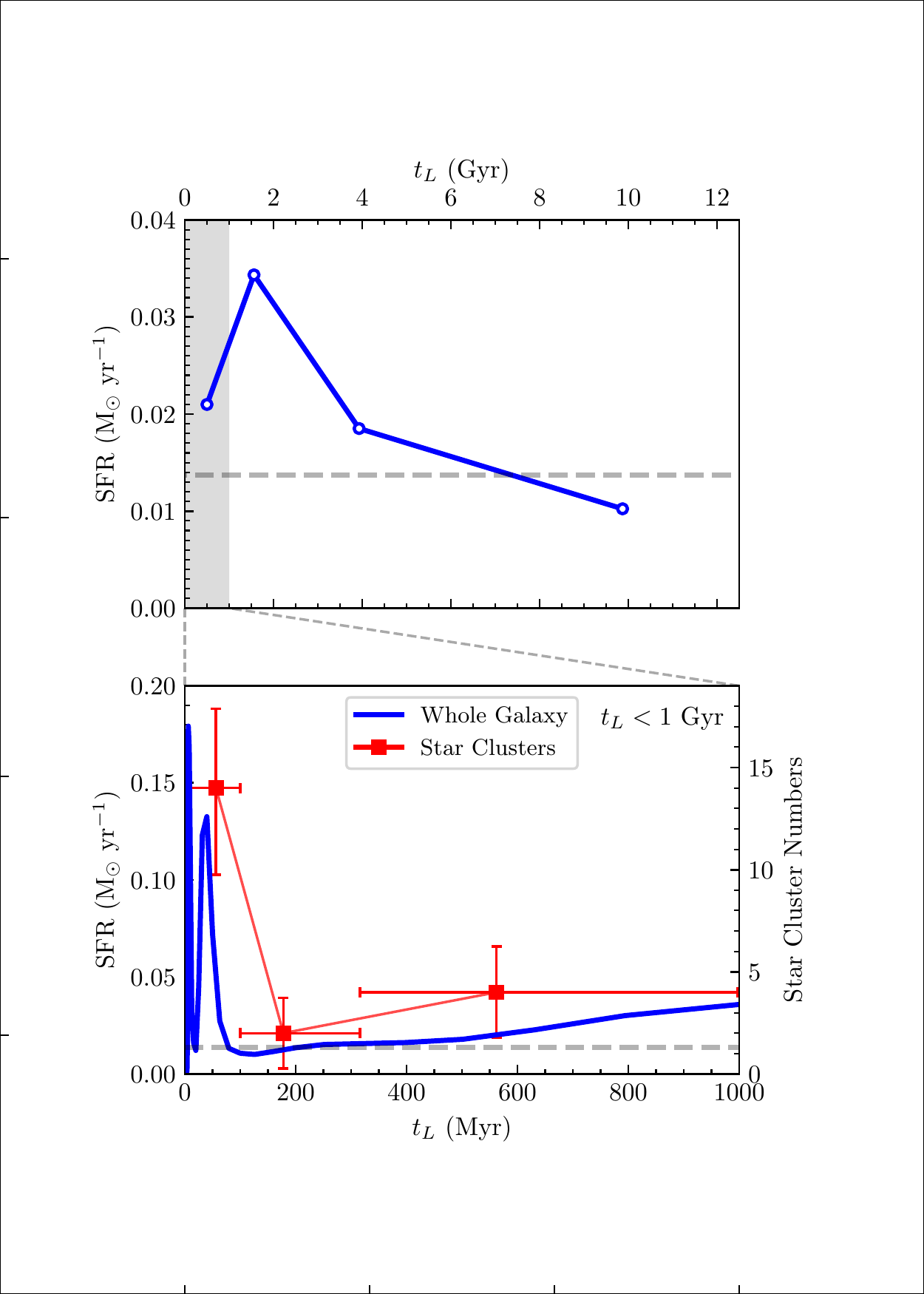}
\caption{Integrated star formation history of the whole galaxy ({\it blue}) and the age distribution of star clusters ({\it red}, bottom panel). The bottom panel is an enlarged view of the last 1 Gyr. The leftmost data point of the top panel represents the average star formation rate of the last 1 Gyr, and the other three points correspond to the star formation rate averaged over logarithmic age intervals of 0.4 dex. In the lower panel, the red horizontal error bars correspond to the (logarithmic) age intervals used for cluster counting, while the red vertical error bars account for the measurement uncertainties of ages and counting noise.}
\label{fig:sfh_sc}
\end{figure}

The global star formation history of NGC 2915 is derived by integrating the spatially resolved star formation histories and shown as blue curves in Figure \ref{fig:sfh_sc}. In addition, number counts of clusters in logarithmic age intervals are overplotted as red curves in the lower panel of Figure \ref{fig:sfh_sc}. As a compromise between sample size, completeness and lookback time, here we only consider clusters more massive than the 90\% completeness limit at the age of $10^{9}$ yr ($\sim1.25\times10^4M_{\odot}$, Figure \ref{fig:cluster_m_age}).

Theoretical studies suggest that cluster dissolution or fading may result in a approximately uniform $\log$(Age) distribution of clusters for a constant cluster formation history \citep[e.g.,][]{Elmegreen2010}, and therefore deviations of the observation from a uniform distribution may indicate a nonconstant cluster formation history. It is encouraging to see that the overall trend of cluster distribution roughly matches that of the star formation history derived from full-spectrum fitting for NGC 2915. Cluster distributions have been used to probe the star formation histories of interacting galaxies in the literature \citep[e.g.,][]{Bastian2005,Zhang2020}. Nevertheless, it is unclear to what extent (and how) the observed cluster distributions have been affected by disruption or fading effects, so it is not straightforward to convert the observed number counts to cluster formation rates.

The global star formation history peaked around $\sim$ 2 Gyr ago, which is also when the most massive cluster in NGC 2915 formed (Figure \ref{fig:cluster_m_age}). In the last $\sim$ 1 Gyr, the global star formation rate has been declining and $\sim$ 10\% of the total stellar mass was formed. The most recent episode of active star formation reached its maximum around $\sim$ 50 Myr ago (lower panel of Figure \ref{fig:sfh_sc}), with a duration of $\sim$ 50 Myr, and has formed $\simeq$ 3\% of the total stellar mass. The cluster age distribution also peaks at $\sim$ 50 Myr or so, in broad agreement with the recent star formation history.

It is conceivable that episodes of bursty star formation like the most recent one may have recurred in the past, but the coarser time resolution at older ages precludes us from recovering them. Nevertheless, we can estimate that such bursty star formation events have recurred less than 4 times (10\%/3\%) in the past 1 Gyr. Our spatially resolved star formation histories and the distribution of star clusters (Section \ref{sec:sc_history}) imply that recurrence of bursty star formation episodes occurred in different spatial locations.

\subsection{Excitation sources of the gaseous nebulae: no evidence for an active galactic nucleus}

\begin{figure*}
\centering
\includegraphics[width=1\textwidth]{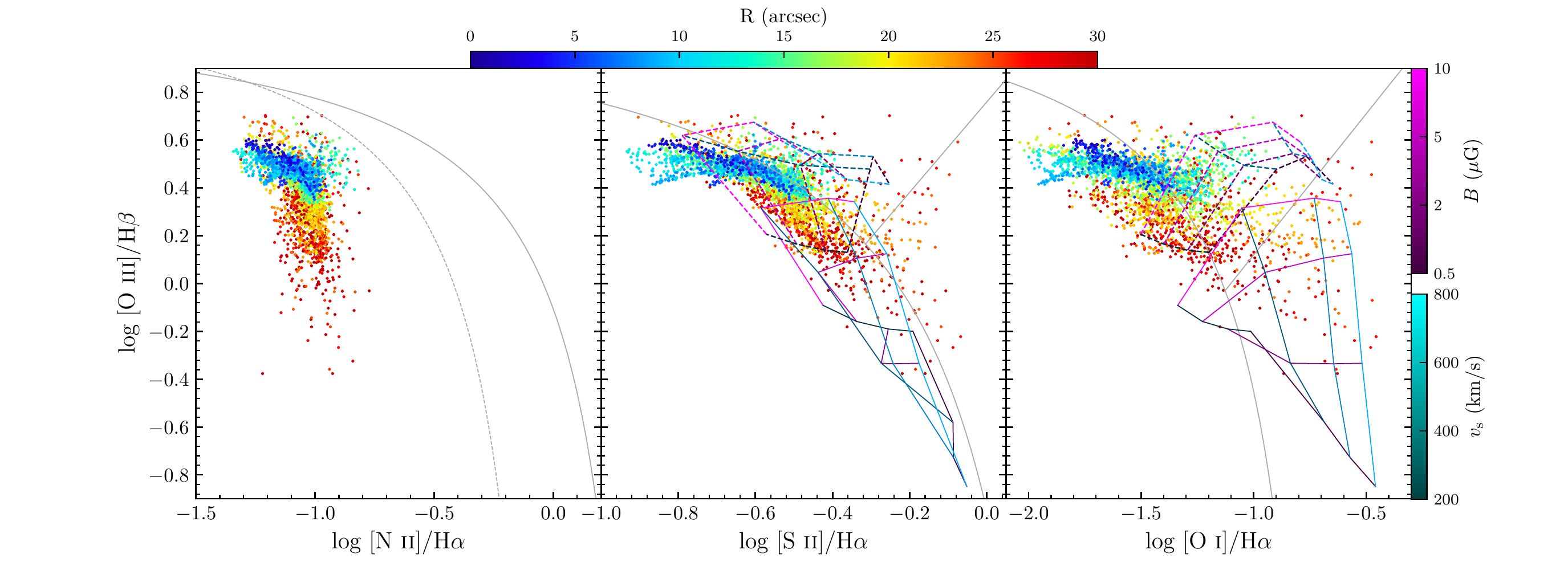}
\caption{The \cite{Veilleux1987} line-ratio diagnostic diagrams of NGC 2915. Each point represents one spaxel and color-coded by the galactocentric radius. The dashed curve in the first panel is the boundary between empirical pure H {\sc ii} region emission and ``composite'' emission proposed by \citet{Kauffmann2003}. The solid curves in the three panels are the \cite{Kewley2001} maximum starburst line. The gray straight lines in the middle and right panels are the empirical boundary between Seyferts and LI(N)ERs proposed by \citet{Kewley2006}. The model grids are the MAPPINGS III shock (solid grids) and shock+precursor (dashed grids) models from \citep{Allen2008} for a SMC-like metallicity. The grid lines are color-coded according to shock velocity $v_s$ and magnetic field $B$.}
\label{fig:bpt}
\end{figure*}

\begin{figure*}
\centering
\includegraphics[width=1\textwidth]{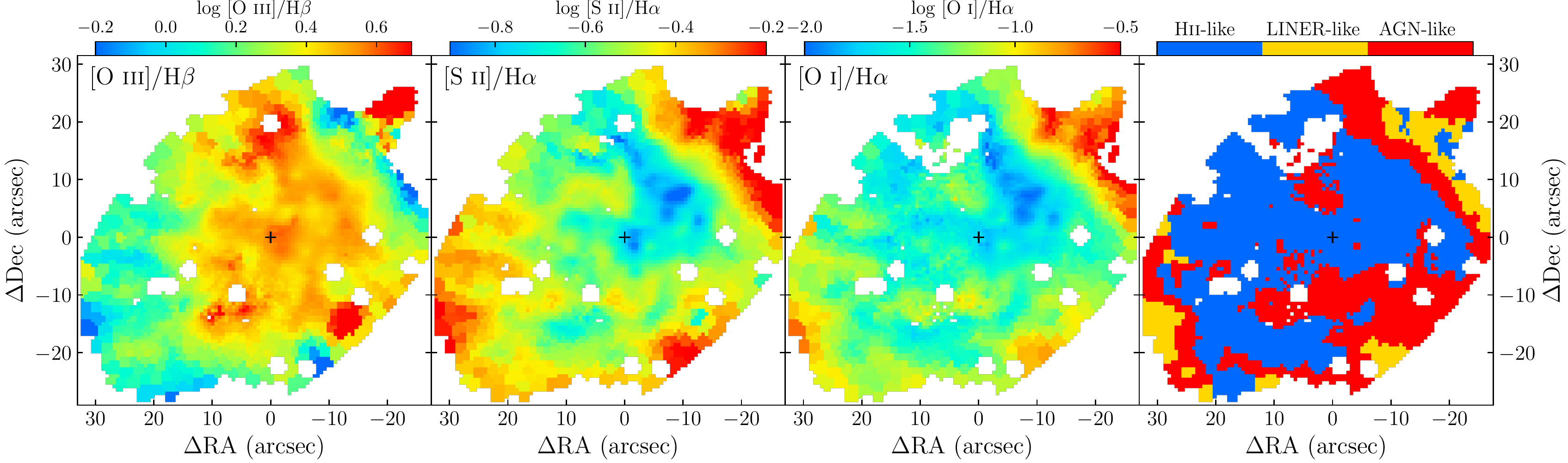}
\caption{Spatial variation of the emission line flux ratio [O {\sc iii}]$\lambda 5007$/H$\beta$, [S {\sc ii}]$\lambda \lambda 6717,6731$/H$\alpha$ and [O {\sc i}]$\lambda 6300$/H$\alpha$. Only spaxels with ${\rm S/N}>10$ in the involved emission lines are plotted. The rightmost panel indicates the spatial distribution of possible excitation sources implied by the [O {\sc iii}]$\lambda 5007$/H$\beta$ vs. [O {\sc i}]$\lambda 6300$/H$\alpha$ diagnostic diagram.}
\label{fig:lineratio}
\end{figure*} 

In order to explore the excitation mechanisms of the gaseous nebulae in NGC 2915, we turn to the three commonly used line-ratio diagnostic diagrams \citep{Veilleux1987}, involving [O {\sc iii}]/H$\beta$, [N {\sc ii}]/H$\alpha$, [S {\sc ii}]/H$\alpha$, and [O {\sc i}]/H$\alpha$. The [O {\sc iii}]/H$\beta$ versus [N {\sc ii}]/H$\alpha$ diagram is known as the BPT diagram \citep{Baldwin1981}, which is one of the most important diagnostics for distinguishing the excitation by star formation and active galactic nuclei (AGN). [S {\sc ii}] and [O {\sc i}] lines are produced by partially ionized zones of gaseous nebulae and are not as sensitive to metallicities as [N {\sc ii}], which makes them ideal tracers of shocks and weak or low-metallicity AGNs.

Figure \ref{fig:bpt} shows the distribution of the Voronoi-binned spaxels of NGC 2915 on the three diagnostic diagrams. The data points are color-coded by galactocentric radius and only spaxels with ${\rm S/N}>10$ in all the six involved emission lines are plotted. Almost all of the spaxels in NGC 2915 are well below the theoretical ``maximum starburst line'' \citep{Kewley2001} and are also below the \cite{Kauffmann2003} empirical ``purely star-forming line'' on the BPT diagram. On the other two line-ratio diagrams, a fraction of spaxels fall into the regime of AGN- or LINER-like excitation, and the fraction is higher in the [O {\sc iii}]/H$\beta$ versus [O {\sc i}]/H$\alpha$ diagram. All but several unconnected spaxels with AGN- or LINER-like spaxels are distributed at large galactocentric distances (rightmost panel of Figure \ref{fig:lineratio}), which basically rules out AGN as a possible excitation source.

We also make an attempt to search for the forbidden high-ionization coronal lines (e.g., [Fe VII]$\lambda$6087, [Fe X]$\lambda$6374, [Fe XI]$\lambda$7892) that have ionization potentials $\gtrsim$ 100 eV \citep[e.g.,][]{Gelbord2009, Negus21}. Detection of these emission lines is considered to be a signature of AGN activity \citep[e.g.,][]{Molina2021}. No such emission lines with ${\rm S/N}>3$ are detected across the observed field, again disfavoring the presence of an AGN.

While there is no evidence for an AGN, excitation sources other than pure star formation, such as shocks, are still possible. In the two diagnostic diagrams involving [S {\sc ii}] and [O {\sc i}] lines, the \cite{Allen2008} shock and shock+precursor models (with one-fifth of solar metallicity) overplotted in the middle and right panels of Figure \ref{fig:bpt} appear to explain the excitation of spaxels above the maximum starburst line to a great extent. Figure \ref{fig:lineratio} further shows the line-ratio maps, where the highest values of the shock-sensitive [S {\sc ii}]/H$\alpha$ and [O {\sc i}]/H$\alpha$ are found at the outer edges of the observed field. Similar findings are also reported in several other dwarf galaxies \citep{Fensch2016,Cairos2017a,Bik2018}. Nevertheless, an inspection of the emission line velocity dispersion map (Figure \ref{fig:vsigma}) reveals that the regions with the highest [S {\sc ii}]/H$\alpha$ and [O {\sc i}]/H$\alpha$ tend to have relatively low velocity dispersion ($\lesssim$ 30 km/s), which raises doubt about their shock-excited nature and instead favors diffuse ionized gas (possibly) ionized by photons leaking from H {\sc ii} regions.

\section{Summary}\label{sec:summary}

In this work, we used the deep integral field spectroscopic data from the VLT/MUSE instrument to investigate the kinematics, stellar populations, and excitation sources of the blue compact dwarf galaxy NGC 2915. The MUSE observations cover the inner exponential disk of the galaxy. In particular, we performed a joint fitting of the MUSE spectra and HST broadband photometry to obtain the spatially resolved star formation histories, studied the spatial distribution and stellar populations of bright star clusters ($V < 22.8$ mag; $M_{V} < -5.3$ mag) detected based on the MUSE data, compared the stellar and gaseous kinematics, and explored the excitation sources of the gaseous nebulae based on the classical line-ratio diagnostic diagrams. Our main results are summarized below.

\begin{itemize}

\item[1.] The stellar disk exhibits a relatively weak but significant rotation within the central $\sim$ 0.5 kpc in radius. The stellar rotation axis appears to be anti-parallel to that of the extended neutral H {\sc i} disk (Section \ref{sec:velfield}). The kinematics of the ionized gas is significantly disturbed either by stellar feedback associated with the active star formation or gas infall/inflow. In contrast to the ionized gas, the neutral H {\sc i} gas does not appear to deviate significantly from regular rotation (e.g., Figure \ref{fig:deltavmap}), implying that the two gas phases are largely spatially decoupled along the line of sight.

\item[2.] The global star formation history peaked around a couple of gigayears ago (Section \ref{sec:sfh}), when the most massive star cluster was formed. The most recent episode of bursty star formation happened around $\sim$ 50 Myr ago and has lasted for $\sim$ 50-100 Myr. It is intriguing to note that the observed cluster age distribution roughly follows the trend of star formation histories. This recent episode of bursty star formation has formed $\sim$ 3\% of the total stellar mass. Episodes of bursty star formation similar to the most recent one may have recurred less than 4 times in different locations (but largely confined within the central 0.4 kpc in radius) in the past 1 Gyr.

\item[3.] The average stellar metallicities have a nearly flat radial profile in the central 0.4 kpc in radius, beyond which the metallicities drop steeply with radius. In contrast to the stellar metallicities, the average gas-phase metallicities have a positive radial gradient ($\sim$ 0.14 dex/kpc), which is unexpected for a system whose metal production, mixing, and loss are in equilibrium. This confirms the finding of \cite{Werk2010} based on spectroscopy of five H {\sc ii} regions. We found evidence for both metal-rich gas outflow and metal-poor gas infall or inflow that may collectively contribute to the positive gas-phase metallicity gradient. In particular, the metal-poor gas infall is unveiled by both signatures of cloud--cloud collisions (Section \ref{sec:disturbedgas}). An unusually small difference between the gaseous and stellar metallicities (Section \ref{sec:csfr}; Figure \ref{fig:age_metal_profile}) also points to either significant inflow or outflow.

\item[4.] Based on the classical line-ratio diagnostic diagrams of \cite{Veilleux1987}, no evidence for an AGN is found in NGC 2915 (Figures \ref{fig:bpt} and \ref{fig:lineratio}). Regions outside the active star-forming area tend to have higher [S {\sc ii}]/H$\alpha$ and [O {\sc i}]/H$\alpha$ line ratios but lower gas velocity dispersions ($\lesssim$ 30 km/s) than the central regions, suggesting a dominant contribution from diffuse ionized gas. 

\end{itemize}

The analysis presented in this work points to a highly complex inner disk of NGC 2915 in terms of the relative distribution of stars and gas. Indeed, \cite{Elson2010} found that the complex H {\sc i} gas distribution of the inner $\sim$ 2 kpc cannot be explained by either pure circular rotation (even allowing for radial variation of disk inclinations) or highly turbulent velocity dispersion. There are at least two nonaxisymmetric H {\sc i} features in the central $\sim$ 2-3 kpc, one is a stream-like feature in the southeast along the major axis that is visible in the velocity field and causes an abrupt decline in radial velocities (Section \ref{sec:velfield}), and the other is a plume-like feature in the northwest also along the major axis that is visible in the column density and moment-2 map \citep{Elson2010}. These features demonstrate an ongoing tidal interaction or gaseous accretion. The discovery of a counter-rotating central stellar disk implies an external origin of the gaseous material that has sustained the recent episodes of bursty star formation.

\begin{acknowledgements}

We thank the anonymous referee for their comments on the manuscript. We thank E. C. Elson for sharing his H {\sc i} data. We acknowledge support from the China Manned Space Project (Nos. CMS-CSST-2021-A07, CMS-CSST-2021-B02), the NSFC grant (Nos. 11421303, 11973038, 11973039, 12122303, 12233008), and the CAS Pioneer Hundred Talents Program, the Strategic Priority Research Program of Chinese Academy of Sciences (Grant No. XDB 41000000) and the Cyrus Chun Ying Tang Foundations. Z.S.L. acknowledges the support from China Postdoctoral Science Foundation (2021M700137).

\end{acknowledgements}

\bibliography{reference}
\bibliographystyle{aa}

\end{document}